# On Nernst effect and quantum oscillations in NCCO

## Partha Goswami

D.B.College, University of Delhi, Kalkaji, New Delhi-110019, India.

**Abstract.** We start with a mean field Hamiltonian in momentum space for NCCO to investigate the magnetic quantum oscillations (MQO) exhibited by this system in the normal state. We first obtain an expression for the entropy of the system which is useful to explain the Nernst effect. We next derive the density of states at Fermi energy and the longitudinal electrical conductivity expression to explore the possibility of MQO. We find that the slow (1/B)-oscillations, with frequency nearly 330 T, arise from the hole pockets at the optimal doping. The faster oscillations at slightly higher doping level is found to have the frequency nearly equal to 900 T.

## 1 Introduction

In this communication, starting with the mean field Hamiltonian in momentum space of the type proposed by Yuan et al. [1] for the electron-doped superconductor $Nd_{2-x}Ce_xCuO_4$(NCCO), we show that the slow (1/B)-magnetic quantum oscillations are possible for the optimally doped sample(x = 0.15) at frequency ~330 T in the Neel ordered state. The oscillation is found have its origin in the minority carrier pocket of the reconstructed Fermi-Luttinger surface obtained using the 'maximal gradient method' [2]. The faster oscillations at the doping level x = 0.17 is found to have the frequency nearly equal to 900 T. We also report that the large Nernst signal is likely to arise from the Fermi surface (FS) patches at $\mathbf{k} = (\pm\pi/2, \pm\pi/2)$ and the boundary of the Fermi pockets around $\mathbf{k} = [(\pm\pi, 0), (0,\pm\pi)]$. At certain points ('Hot spots') on the boundary of the magnetic Brillouin zone(MBZ) close to the Fermi pockets we have found discontinuities in the entropy density difference (EDD) between the anti-ferromagnetic state and the normal state; on the boundaries of the electron and hole pockets EDD peak dramatically.

The paper is organized as follows: In section 2 we outline the FS reconstruction exercise and the derivation of the momentum dependent Nernst signal. In section 3 we examine the quantum oscillations in electrical conductivity in the presence of changing magnetic field. We conclude the paper in section 4 presenting a brief summary and remarks related to the investigations carried out.

## 2 Mean field treatment and Nernst signal

### 2.1 Mean-field treatment

In the second-quantized notation, for the anti-ferromagnetic state of NCCO, the mean-field Hamiltonian of the type proposed by Yuan et al.[1] by considering two sub-lattices D and E with corresponding operators d and e, respectively, can be expressed as $H = 2NJ(\chi^2 + m^2) + \sum_{k\sigma} \Phi^\dagger_{k,\sigma} E(k) \Phi_{k,\sigma}$ where $\Phi^\dagger_{k,\sigma} = (d^\dagger_{k,\sigma}\ e^\dagger_{k,\sigma})$ and

$$E(k) = \begin{pmatrix} \varepsilon_2(k) - \mu - 2Jm\sigma & \varepsilon_1(k) \\ \varepsilon_1(k) & \varepsilon_2(k) - \mu + 2Jm\sigma \end{pmatrix}. \qquad (1)$$

The dispersions ($\varepsilon_1(k)$, $\varepsilon_2(k)$), involving t, t´, t´´ which are the hopping elements between the nearest, next-nearest(NN) and NNN neighbours respectively, are given by

$$\varepsilon_1(k) = (-2tx - J\chi)(\cos k_x a + \cos k_y a),$$

$$\mathcal{E}_2(k) = -4t'x \cos k_x a \cos k_y a - 2t''x(\cos 2k_x a + \cos 2k_y a). \tag{2}$$

The wave vectors **k** are restricted to the magnetic Brillouin zone (MBZ). The hopping integrals (t, t'') < 0 whereas t' > 0. The quantity χ corresponds to the uniform fermionic bond order, and 'm' to the anti-ferromagnetic (AF) order parameter. For the mean-field treatment under consideration, χ and m could be determined by the minimization of the Free energy. The quantity x denotes the doping level and N is the total number of sites. The chemical potential μ is obtained applying Luttinger theorem.

The eigenvalues of the matrix in (1) are $E_k^{(r=U,L)} = (\mathcal{E}_2(k) - \mu \pm w_k)$ where $w_k = (\mathcal{E}_1(k)^2 + 4J^2 m^2)^{1/2}$. With these eigenvalues we find that, for the D sub-lattice, the momentum dependent occupancy (MDO) is

$$f_D(k) = \sum_\sigma [u_{k\sigma}^2 (\exp(\beta E_k^{(U)})+1)^{-1} + v_{k\sigma}^2 (\exp(\beta E_k^{(L)})+1)^{-1}] \tag{3}$$

and the corresponding spectral weight at non-zero frequency $A_D(k,\omega) = (2\pi/|\omega|) \sum_\sigma [u_{k\sigma}^2 \delta(1 - E_k^{(U)}/\omega) + v_{k\sigma}^2 \delta(1 - E_k^{(L)}/\omega)]$ where $u_{k\sigma}^2 = (1/2)[1 - (2Jm\sigma / w_k)]$ and $v_{k\sigma}^2 = (1/2)[1 + (2Jm\sigma / w_k)]$. The relative excess of the fermions of one spin type, for the sub-lattice D(E), is given by $R_{D(E)} = (N)^{-1} \sum_k r_{D(E)}(k)$ where $r_{D(E)}(k) = (f_{D(E)}(k,\uparrow) - f_{D(E)}(k,\downarrow))$. We find that $r_D(k) = -r_E(k)$ as it should be. The density of states (DOS) is given by $\rho_D(k,\omega) = A_D(k,\omega) / 2\pi$. The MDO for the E sub-lattice, on the other hand, is given by $f_E(k) = \sum_\sigma [v_{k\sigma}^2 (\exp(\beta E_k^{(U)})+1)^{-1} + u_{k\sigma}^2 (\exp(\beta E_k^{(L)})+1)^{-1}]$. Note that $f_D(k) = f_E(k)$ and $A_D(k,\omega) = A_E(k,\omega)$. The values of the parameters, as in ref.[1] for 15% doping, are chosen to be |t|= 1, t' = 0.32, t'' = −0.16, J = 0.3, χ = −0.189, m = 0.144, and μ = −0.00385 for the numerical calculations below. In Fig.1 we have shown the plot of the MDO, f(k), on the Brillouin zone for this doping level. The MDO is unity all over except at the humps (electron regions) centered around **k** = (±π/2, ±π/2), where it is greater than unity, and pits (hole pockets) around **k** = [(±π, 0), (0,±π)] (excluding their boundaries) where it is close to zero. Since with the finite electron doping level the occupancy corresponding to the electron region can be expected to be greater than one, the reason for the identification made above is obvious.

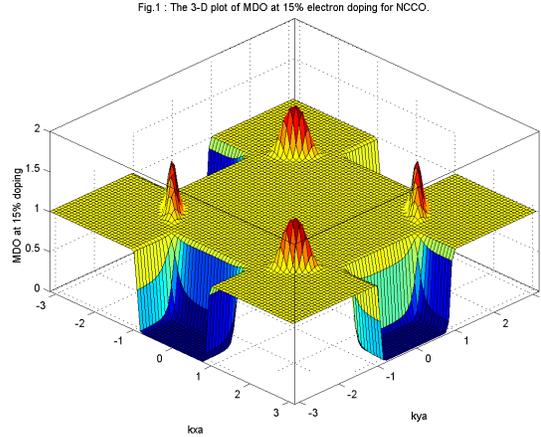

**Figure 1** The 3-D plot of the MDO shows that it is unity all over except at the humps centered around k = (±π/2, ±π/2), where it is greater than unity, and pits around k = [(±π, 0), (0,±π)] where it is close to zero. The humps correspond to electron pockets and the pits to hole pockets on the Brillouin zone.

The DOS (contour) plot at ω = −μ, in Fig. 2(a), apart from patches around **k** = (±π/2, ±π/2), show the boundary of the Fermi pockets around [(±π, 0), (0,±π)]. These pockets are the regions where DOS is suppressed. Since we are interested in the magnetic oscillations and not the ARPES outcomes, the zero frequency spectral weight has not been considered. We have obtained the Fermi-Luttinger surface (FLS)(see Fig.2(b)), with small electron and hole pockets, using the expression for f(k) = $\sum_\sigma f_D(k,\sigma) = \sum_\sigma f_E(k,\sigma)$ and applying the 'maximal gradient method'[2]. The method is based on the fact that the FLS is given by the set of **k**-values for which the momentum dependent occupancy, f(k), shows a jump discontinuity in the zero temperature limit. When this discontinuity is smeared out, say, by thermal broadening, the gradient of f(k), $|\nabla f(k)|$, is assumed to be maximal at the locus of the underlying FLS. In particular, we find the existence of the Fermi pockets around $\mathbf{k} = [(\pm\pi, 0), (0,\pm\pi)]$, and FS patches around $\mathbf{k} = (\pm\pi/2, \pm\pi/2)$. These features are consistent with the experimental results found in ref.[3].

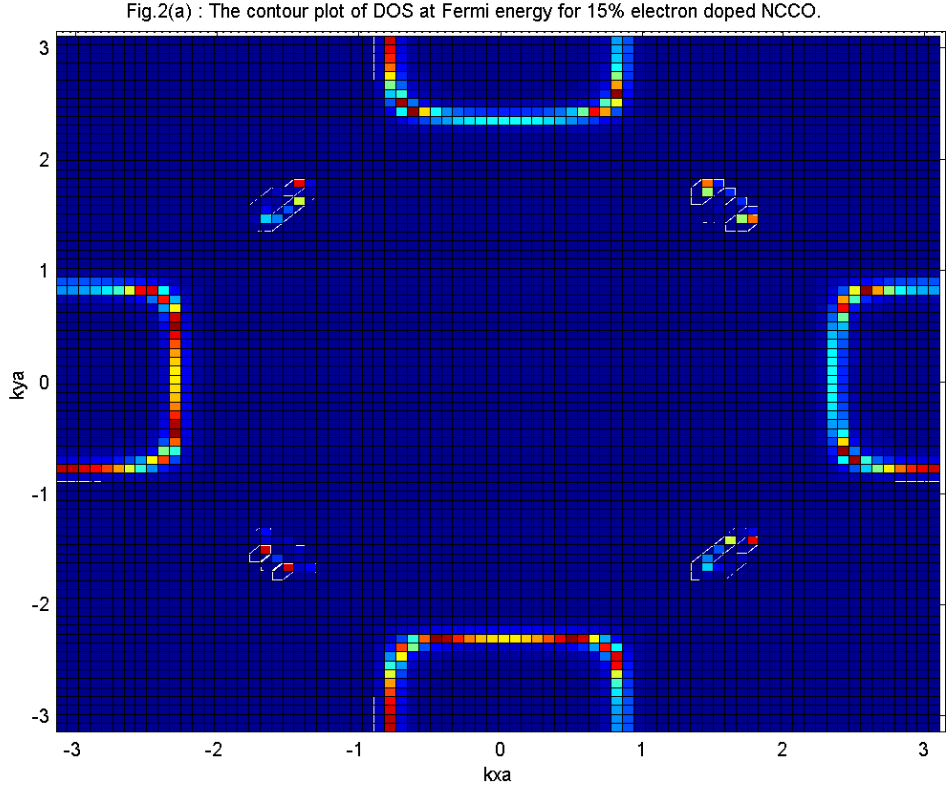

Fig.2(a) : The contour plot of DOS at Fermi energy for 15% electron doped NCCO.

In the plot of the FLS, the sign of the first neighbor hopping term 't' is not important. However, the signs of both the first neighbor and the third neighbor hopping terms have to be the same and the sign of the second neighbor term different. Then and only then the distinct Fermi pockets appear.

### 2.2 Nernst signal

We find that the thermodynamic potential per site is given by $\Omega = \Omega_0 - 2(\beta N)^{-1} \sum_{k,r(=U,L)} \{\ln\cosh(\beta E_k^{(r)}/2)\}$ where $\Omega_0 = 2J(\chi^2+m^2)+(N)^{-1}\sum_{k,r(=U,L)} E_k^{(r)}$. The dimensionless entropy per site is given by $S = \beta^2 (\partial\Omega/\partial\beta)$. Assuming that $(m,\chi)$ are independent of temperature, we obtain for the AF phase ($m \neq 0$) and the paramagnetic normal phase ($m=0$) $S_{AF} = (2/N) \sum_{k,r(=U,L)} s_{AF}(\mathbf{k})$ and $S_N = (2/N)\sum_{k,i} s_N(\mathbf{k})$ where

$$s_{AF}(\mathbf{k}) = [\ln(1/2) + \ln(1+\exp(-\beta E_k^{(r)})) + (\beta E_k^{(r)} + \beta^2 (\partial E_k^{(r)}/\partial\beta))(\exp(\beta E_k^{(r)})+1)^{-1}], \qquad (4)$$

$$s_N(\mathbf{k}) = [\ln(1/2) + \ln(1+\exp(-\beta \mathcal{E}_i(k))) + (\beta \mathcal{E}_i(k) + \beta^2 (\partial \mathcal{E}_i(k)/\partial\beta))(\exp(\beta \mathcal{E}_i(k)+1)^{-1}] \qquad (5)$$

where $\mathcal{E}_i(k) = (\mathcal{E}_2(k) \pm \mathcal{E}_1(k))$. The entropy-expression is useful to explain Nernst effect. The thermoelectricity as a probe of the normal state of NCCO is still largely unexplored. In what follows we discuss this issue briefly: When an electric field $\mathbf{E}$ is applied within the conducting plane of the system and a magnetic field $\mathbf{B}$ in the perpendicular direction, a quasi-particle in normal state drifts with velocity $\mathbf{v}_D$ perpendicular to both **B** and **E** ($\mathbf{v}_D = (\mathbf{E} \times \mathbf{B})/B^2$). The heat current parallel to $\mathbf{v}_D$ is then given by $\mathbf{J}_H = \beta^{-1} S_{total} \mathbf{v}_D$, where $S_{total}$ is the total entropy associated with the quasi-particles in the normal state. In fact, at

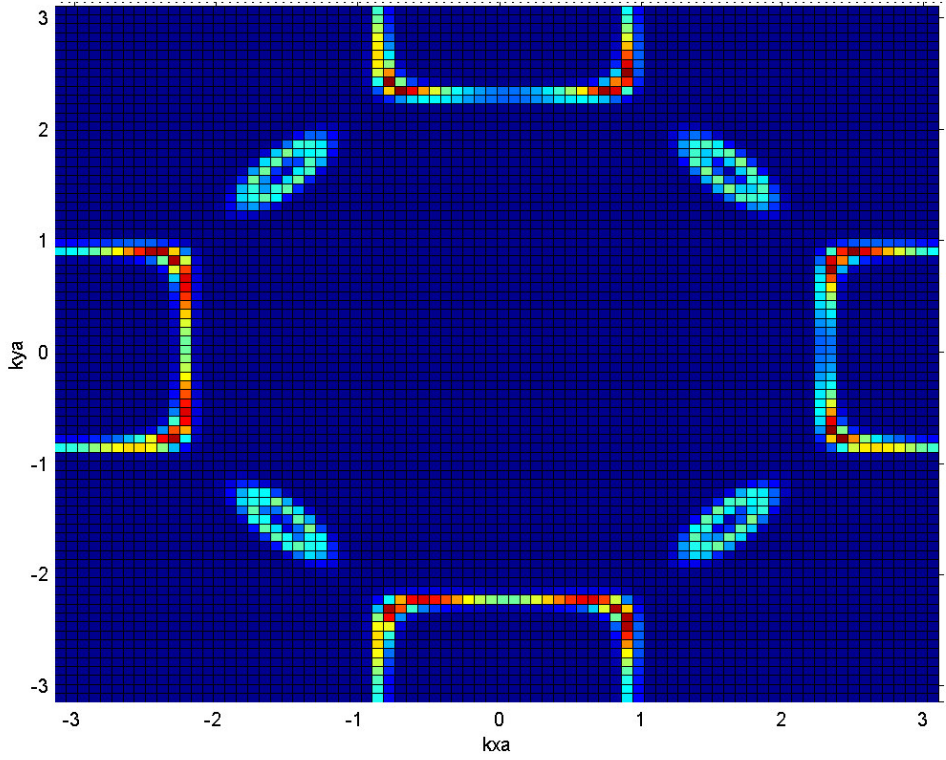

Fig.2(b) : The contour plot of reconstructed Fermi surface by the maximal gradient method at 15% electron doping for NCCO.

**Figure 2.** (a) The contour plot of density of states at $\omega = -\mu$ in the absence of magnetic field. It shows the FS patches around $\mathbf{k} = (\pm\pi/2, \pm\pi/2)$ as well as the boundary of the Fermi pockets around $[(\pm\pi, 0), (0,\pm\pi)]$.. (b) The contour plot in this figure shows the outcome of the Fermi surface reconstruction exercise carried out applying the 'maximal gradient method'(see ref.[2]). The patches and the pockets mentioned above duly appear around $\mathbf{k} = (\pm\pi/2, \pm\pi/2)$ and $\mathbf{k} = [(\pm\pi, 0), (0,\pm\pi)]$ respectively.

$B = 0$, the electric field tends to become orthogonal to the thermal current. The presence of magnetic field takes away this alignment. The quantity $S_{total} = S_{Landau} + S_{AF}$ where $S_{Landau} = \sum_n [\ln(1+\exp(-\beta E_n)) + \beta E_n (\exp(\beta E_n)+1)^{-1}]$, $E_n = (n+(1/2))\hbar\omega_c$ (where n = 0,1,2,...), and $\omega_c = eB/m^*$. The sum in $S_{Landau}$ has to be taken over all the Landau levels. For small T and large B, this is well approximated by taking the n = 0 and n = 1 Landau levels. The entropy $S_{AF}$ is obtained as in Eq.(4). The Nernst effect corresponds to the off diagonal component of the thermoelectric power in the presence of magnetic field. In the configuration discussed above, the Nernst coefficient ($S_{xy} = -[(k_B\rho/B) \sum_k s_{total}(\mathbf{k})]$) calculation appears to be possible. Here $\rho$ is the magneto-resistivity within the two level approximation. For B ~ 10 T, $\rho$ ~ 70 mΩ-cm. In Fig.3 we have shown a 3-D plot of the total entropy density $s_{total}(\mathbf{k})$ on the Brillouin zone (for B ~ 10 T) for the 15% doped sample. It is clear from the figure that the large Nernst signal is likely to arise from the FS patches at $\mathbf{k} = (\pm\pi/2, \pm\pi/2)$ and the boundary of the Fermi pockets around $\mathbf{k} = [(\pm\pi, 0), (0,\pm\pi)]$. The existence of an AF order related gap lowers the carrier density as the gap destroys much of the Fermi surface. This restricts the phase space and leads to an increase in the mean free-path of the quasi-particles. Possibly, these effects conspire in such a way so as to create a large Nernst signal. We find that discontinuities in the entropy density difference $\Delta s(\mathbf{k}) = (s_{AF}(\mathbf{k}) - s_N(\mathbf{k}))$ between the Neel ordered state and the normal state occur at the eight points (Co-ordinates ~ ($\pm 1.0, \pm 2.1$),($\pm 2.1, \pm 1.0$)), referred to as the

**Table1.** The data in the table indicates that the discontinuity in $\Delta s(\mathbf{k})$ occurs at the point ~(2.1,1.0) on the boundary of the magnetic Brillouin zone.

| $k_x a$ | 1.7 | 1.9 | 2.1 | 2.3 |
|---|---|---|---|---|
| $k_y a$ | 1.4 | 1.2 | 1.0 | 0.8 |
| $\Delta s(\mathbf{k})$ | 0.6624 | −0.0823 | −0.9501 | 0.6865 |

'Hot spots' located on the boundary of MBZ (see Table 1). On the boundaries of the Fermi pockets and patches, $\Delta s(\mathbf{k})$ peak dramatically.

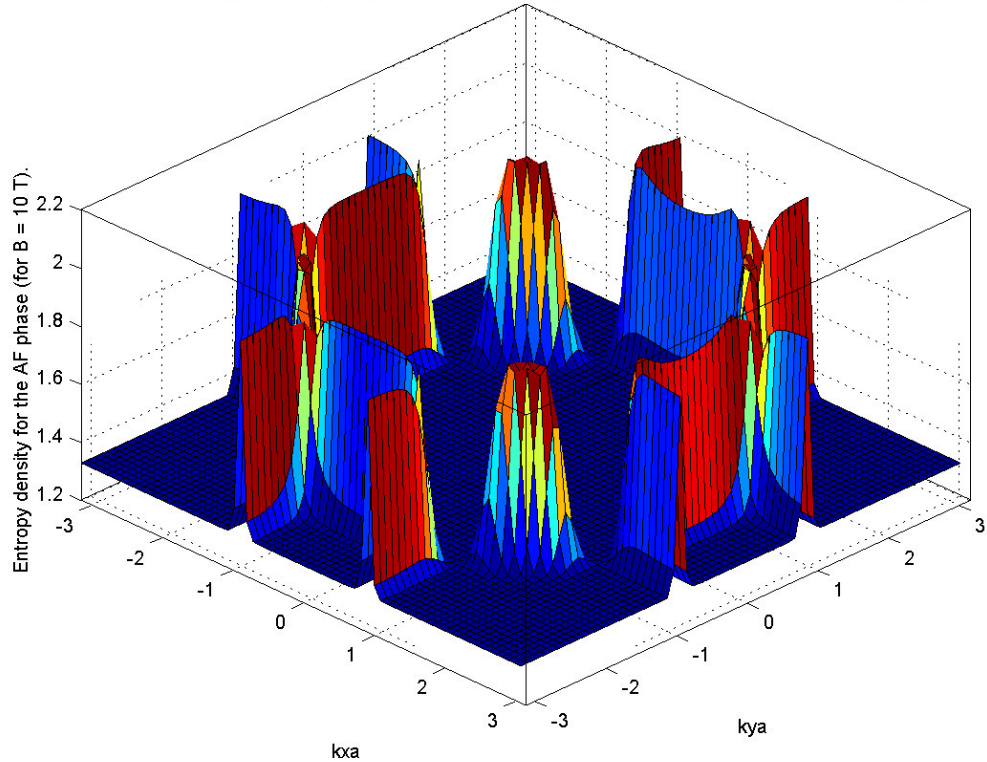

Figure 3 : The 3-D plot of the entropy density (in the presence of B = 10 T) on the Brillouin zone at 15% electron doping for NCCO.

**Figure 3.** Here we have shown a 3-D plot of the total entropy density $s_{total}(\mathbf{k})$ on the Brillouin zone (for B ~ 10 T) for the 15% doped sample. It is clear from the plot that the large Nernst signal is likely

to arise from the FS patches at **k** = (±π/2, ±π/2) and the boundary of the Fermi pockets around **k** = [(±π, 0), (0,±π)].

## 3. Quantum oscillations in conductivity

We shall now discuss the quantum oscillations in conductivity at low temperature in the presence of a magnetic field. For a magnetic field applied in z-direction ( i.e. the vector potential **A** = (0,−Bx, 0) in Landau gauge),starting with the dispersions in Eq. (2) and the renormalized chemical potential $\mu' \equiv \mu - \hbar(\omega_c/2)\sum_{n=0}^{\infty}(2n+1) + (-1)^\sigma (g\mu_B B/2)$ ( where the cyclotron frequenc $\omega_c = eB/m^*$ ($m^*$ is the effective mass of the electrons), and the Zeeman term is ($g\mu_B B/2$)), we write an approximate expression of the spectral function for B ≠ 0 replacing the delta functions in terms of Lorentzians :

$$A(\mathbf{k}, \mu) \sim (2\gamma^2/(\hbar\omega_c)).((|2(E_k^{(U)} - 0.00385)/(\hbar\omega_c)| - \theta)^2 + \gamma^2)^{-1}$$

$$+ (2\gamma^2/(\hbar\omega_c)).((|2(E_k^{(L)} - 0.00385)/(\hbar\omega_c)| - \theta)^2 + \gamma^2)^{-1} \qquad (6)$$

where $\theta = \sum_{n=0}^{\infty}(2n+1)$, $\omega = -\mu$, and $\gamma$ corresponds to the level-broadening due to collisions (this term we have artificially introduced as we have not considered the elastic scattering by impurities in this communication); the Zeeman term is assumed to be insignificant. Alternatively, these delta functions could also be expanded in the Fourier cosine series: $\delta(x-a) = (2\pi)^{-1} + \pi^{-1}\sum_{m=1}^{\infty} \cos[m(x-a)]$. The effective carrier concentration is obtained by the integration $\int d\mathbf{k}(A(\mathbf{k})/2\pi)$ at near zero temperature. Here $\int d\mathbf{k} \rightarrow \int_{-\pi}^{+\pi}(d(k_x a)/2\pi) \int_{-\pi}^{+\pi}(d(k_y a)/2\pi)$. Now with increase in B, the Landau states move to a higher energy, ultimately rising above the Fermi level. They are thereby emptied, and the excess fermions find a place in the next lower Landau level(LL). During the crossing of a LL, its occupation by fermions is halted and then reduced. The electrical conductivity consequently decreases slightly. As the excess fermions get accommodated in the next lower LL, the conductivity rises again. These oscillations in the fermion density in the vicinity of Fermi energy manifest themselves as oscillations in electrical conductivity (Shubnikov-de Haas oscillations (SdHO)). In YBCO these oscillations are believed to have their origin mainly at electron pockets in Fermi surface located around the anti-nodal points. Thus the magnetic field dependent occupancy $n_e([(\pm\pi, 0), (0,\pm\pi)],B,T)$ at these points are expected give a reasonable idea about the frequencies of these oscillations. For NCCO, we are not yet certain about whether the FS patches around $\mathbf{k} = (\pm\pi/2, \pm\pi/2)$ or the pockets around $\mathbf{k} = [(\pm\pi,0),(0,\pm\pi)]$ correspond to the origin of such oscillations. In view of the facts stated above, it is, however, clear that the Fermi energy DOS $(A(\mathbf{k})/2\pi)$ plot, in particular the peak-valley-peak scenario manifestation in Fermi energy DOS, is expected to give an indication of the origin. With this in mind, in Fig.4 we have plotted the quantity $(A(\mathbf{k})/2\pi)$ on the Brillouin zone for nonzero magnetic field B. We have chosen several representative points constituting the set $S_1 = \{(k_x a, k_y a) = (2.40, 0.60), (2.30, 0.00), (2.50, −0.80),……\}$ and the set $S_2 = \{(k_x a, k_y a) = (2.05, 1.30), (1.5,1.5),(1.5,2.0),…..\}$ on the boundaries of the Fermi pocket (around $\mathbf{k} = [(\pm\pi,0),(0,\pm\pi)]$) and the patch (around $\mathbf{k} = (\pm\pi/2, \pm\pi/2)$), respectively. We have next plotted DOS as a function of the magnetic field for these points (see Figs. 5(a) and (b)). In Fig.5(a), which corresponds to the point $(k_x a, k_y a) = (2.40, 0.60)$, we have the peak-valley-peak scenario envisaged above. In fact, for all the points in the set $S_1$ the same thing is observed. In Fig.5(a), the first peak occurs at $B_1 \sim 350$ Tesla while the second one at occurs at $B_2 \sim 1600$ Tesla. In Fig. 5(b) which corresponds to $(k_x a, k_y a) = (2.05, 1.30)$, on the other hand, there is a single peak located roughly at 470 T. The single-peak scenario occurs for all the points in the set $S_2$. These outcomes indicate that for the two-peak case, with the increase in B, the first Landau state after crossing the Fermi level gets emptied and the excess fermions are passed onto the next lower state which, too, crosses the Fermi level at higher magnetic field. However, for the single-peak case, the next state is not able to cross the Fermi level up to the magnetic field as much as 2600 T. It is, thus, clear that the low-frequency (1/B)-oscillations (with frequency ~350 Tesla where the first DOS peak occurs) originates from the boundary of the hole pocket; the fast oscillations may have the contributions from the electron pockets too. Our finding here is broadly in agreement with that of Barnea et al.[4] where they have demonstrated that in an applied external magnetic field the low-energy density of states (DOS) of a fermion system oscillates as a function of energy as well as magnetic field. A proper derivation of the conductivity (analyzing the appropriate two-particle Green's function) is, however, very much required to show the conductivity oscillations clearly. Before we undertake this task, we note that these periodic quantum oscillations are in principle observable in all solid state properties of NCCO in the normal as well as the superconducting states, and have in fact been observed in magneto-resistivity measurements recently[5].

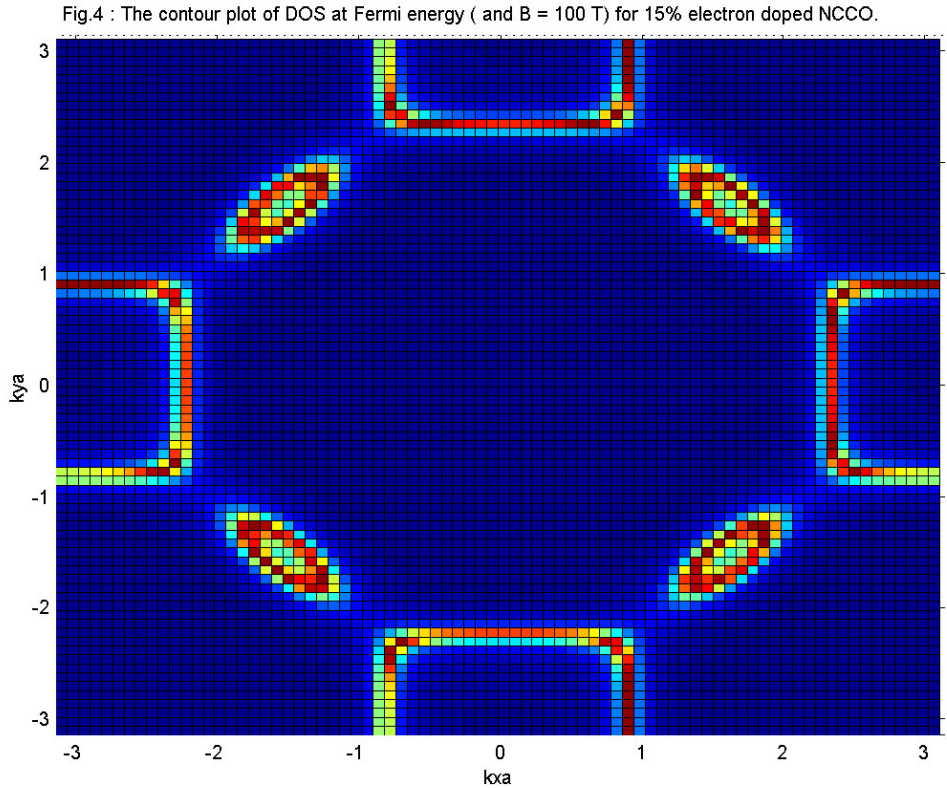

Fig.4 : The contour plot of DOS at Fermi energy ( and B = 100 T) for 15% electron doped NCCO.

**Figure 4** The contour plot of the density of states at Fermi energy (and B = 100 T) for the optimally doped NCCO.

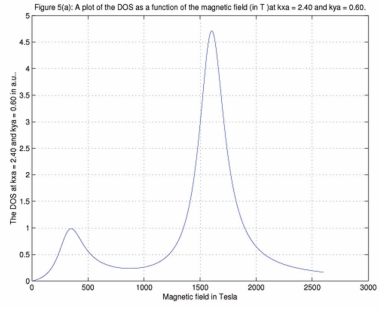
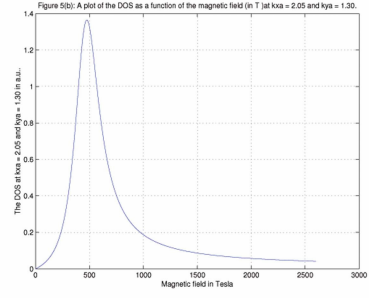

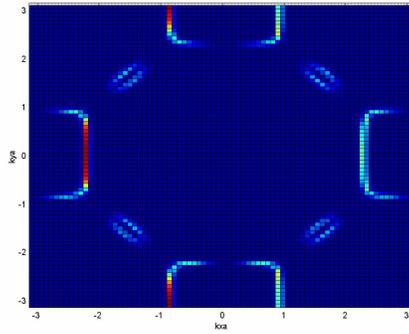
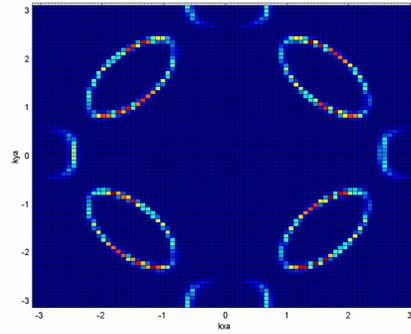

**Figure 5.** The plots of DOS as a function of B for the points (2.40, 0.60) and (2.05, 1.30)). **(a)** The DOS plot corresponding to the point (2.40, 0.60) on the Brillouin zone. **(b)** The DOS plot corresponding to the point (2.05, 1.30) on the Brillouin zone.

**Figure 6.(a)** The contour plot of the conductance density $\sigma'_{xx}(k)$ for the optimally doped (doping level = 0.15) sample, with B = 10 T. The conductance density is found to shoot up at the closed FS patches around **k** = ($\pm\pi/2$, $\pm\pi/2$) and on the boundary of Fermi regions around **k**= [($\pm\pi$,0),(0,$\pm\pi$)]. **(b)** The contour plot of the conductance density for the 17% doped NCCO (for B = 100 T ) is shown in Fig.6(b). The FS patches (or electron pocket) around **k** = ($\pm\pi/2$, $\pm\pi/2$) here are bigger than those of Fig.6(a) and hole pockets are smaller.

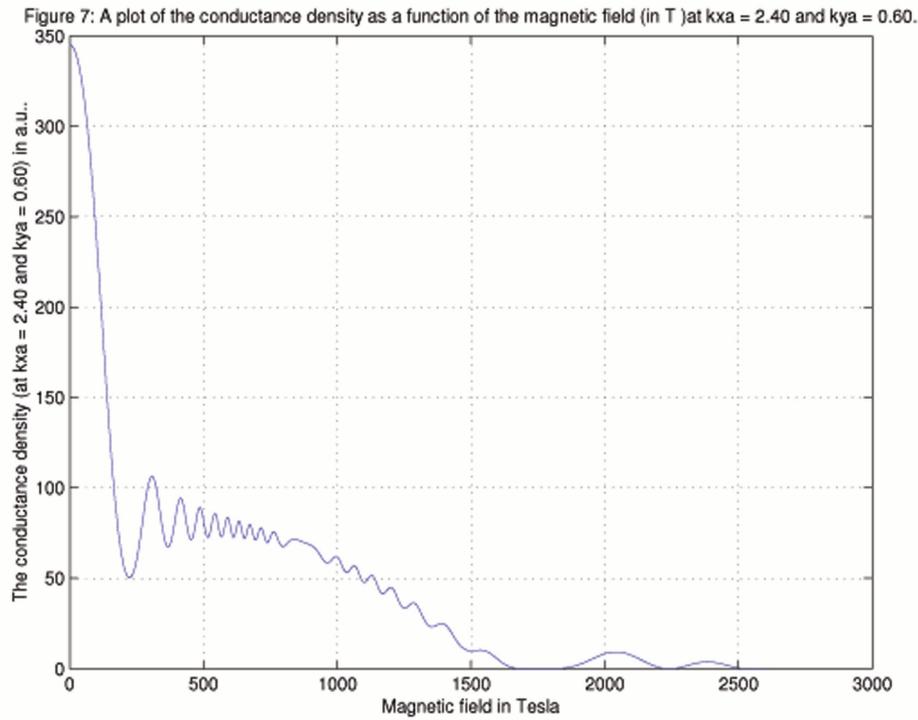

**Figure 7.** A 2-D plot of the conductance density as a function of the magnetic field in Tesla for the point $(k_x a, k_y a) = (2.40, 0.60)$ on the Brillouin zone. In order to obtain this plot the delta functions in the spectral weight have been expanded in the Fourier cosine series: $\delta(x-a) = (2\pi)^{-1} + \pi^{-1}\sum_{m=1}^{\infty} \cos[m(x-a)]$. In this infinite series, 26 terms have been taken into account. The plot is more of illustrative nature than quantitative (see section 4).

We now suppose that a weak time-dependent external force (e.g. electric scalar or vector potential $\mathbf{A}(\mathbf{x},t)$) is applied to the system. Our task is to analyze the current response function $\mathbf{J}(\mathbf{x},t)$ rather than the charge density as the former directly leads to the electrical conductivity. We further assume **A** to represent a transverse field, so that we do not have to worry about the internal field arising from the induced charge densities. It follows [6] that

$$\langle J_\alpha(x,t) \rangle = \langle j_\alpha(x) \rangle + {}_{-\infty}\int^{+\infty} dt' \int dx' \sum_\beta R_{\alpha\beta}(x-x',t-t') \times A_\beta(x',t') \qquad (7)$$

where $R_{\alpha\beta}(x-x',t-t') = -i\,\theta(t-t')\,\langle[j_\alpha(x,t), j_\beta(x',t')]\rangle - (ne^2/m)\,\delta_{\alpha\beta}\,\delta(x-x')\,\delta(t-t')$, n is the carrier concentration and $\alpha,\beta$ are the Cartesian suffixes. The angular brackets denote a thermal average. Equation (7) constitutes what is known as the Kubo formula. The dc conductivity $\sigma_{\alpha\beta}(0)$ is obtained from the Fourier transform $R_{\alpha\beta}(q,\omega)$ in space and time of the retarded response function $R_{\alpha\beta}(x-x',t-t')$: $\sigma_{\alpha\beta}(0) = \mathrm{Lim}_{\omega\to 0, q\to 0}(R_{\alpha\beta}(q,\omega)/i\omega)$. Following the derivation in ref.[6] we find that $\sigma_{\alpha\beta}(0) = \sum_k \sigma'_{\alpha\beta}(k)$ where $\sigma'_{\alpha\beta}(k) \sim -v_\alpha(k)\,v_\beta(k)\int dx A(k,x) A(k,x) (\partial f/\partial x)$, $v_j(k)$ are the velocity functions, $f(x)$ is the Fermi distribution, and $A(k,x)$ is the spectral distribution function of the single-particle Green's function. We have already expressed $A(k,x)$ in the standard Lorentzian form in (6). Upon assuming that $-(\partial f(\varepsilon)/\partial \varepsilon) = \delta(\varepsilon-\mu)$ we find that conductance density $\sigma'_{\alpha\beta}(k) \sim v_\alpha(k)\,v_\beta(k)\,A(k,\mu)\,A(k,\mu)$. The velocity functions for the two bands $E_k^{(r=U,L)} = (\varepsilon_2(k) - \mu \pm w_k)$ are given by $v_j^{(r)}(k) = (\partial E_k^{(r=U,L)}/\partial k_j)$. Through the graphical representations, we now summarize below the results obtained relating to the quantity $\sigma'_{xx}(k)$.

The conductance density $\sigma'_{xx}(k)$ for the optimally doped (doping level = 0.15) sample, with B = 10 T, is found to shoot up at the closed FS patches around $\mathbf{k} = (\pm\pi/2, \pm\pi/2)$ and on the boundary of Fermi regions around **k**= [(±π,0),(0,±π)]. We have shown a representation of this through the contour plot in Fig.6(a). The contour plot of the conductance density for the 17% doped NCCO (for B = 100 T ) is shown in Fig.6(b). The FS patches around **k** = (±π/2, ±π/2) in the latter are bigger than those of the former. It follows from the Onsager relation [6] that the (1/B)-oscillation frequency(nearly equal to 900 T) for the latter will be greater than that of the former. These results are in qualitative agreement with those reported in ref.[4]. We have shown a plot of $σ′_{xx}(k)$, for the optimally doped sample and $(k_x a, k_y a)$ = (2.40, 0.60), as a function of the magnetic field in Fig.7 . The figure clearly shows the conductivity oscillations taking place with the increase in magnetic field. As is noted below Eq.(6), we have artificially introduced a quantity γ which corresponds to the level-broadening due to the collisions. The inclusion of the elastic scattering by impurities is, therefore, required for a comprehensive analysis of this problem.

## 4. Summary and concluding remarks

In summary, we have examined the Nernst effect and quantum oscillation problems for the electron-doped cuprate NCCO by using the model of the type proposed by Yuan et al.[1] at the semi-phenomenological level. The finer constraints of the model, such as the 'no double occupancy is allowed', have strictly not been adhered to hoping that the model essentially is also derivable by a slave-bosonic treatment involving holons and doublons[7]. This task, if accomplished, will provide a microscopic basis to the model. Furthermore, as is evident from Fig.1, our identification of the electron and hole pockets is very much based on the fact that the 'occupancy' can be greater than one. Now, alternatively, we could have taken our starting point as the phenomenological t-t′-t″-U-V model [8] with repulsive on-site U and attractive inter-site V to examine the MQO;. the V–term corresponds to the superconducting( SC)pairing interaction. If this is done, it would be interesting to see how would the findings (with the present starting point) change. For example, the entropy density calculated within the Hartree-Fock approximation framework of the present approach shows that the large Nernst signal is likely to arise from the FS patches at **k** = (±π/2, ±π/2) and the boundary of the Fermi pockets around **k** = [(±π, 0), (0,±π)]. A suitable approximation scheme, such as the Hubbard I, to deal with strong onsite interaction U may not lead to the same result.

As regards quantum oscillations in the conductivity, Figs.5(a) (and (b)) indicate that for the two-peak case, with the increase in B, the first Landau state after crossing the Fermi level gets emptied and the excess fermions are passed onto the next lower state which, too, crosses the Fermi level at higher magnetic field. However, for the single-peak case, the next state is not able to cross the Fermi level up to the magnetic field as much as 2600 T. It is, thus, apparent that the low-frequency (1/B)-oscillations (with frequency ~350 Tesla where the first DOS peak occurs) originates from the boundary of the hole pocket; the fast oscillations may have the contributions from the electron pockets too. Our finding here is broadly in agreement with that of Barnea et al.[4] where they have demonstrated that in an applied external magnetic field the low-energy density of states (DOS) of a fermion system oscillates as a function of energy as well as magnetic field. Moreover, our investigation has led us to a 2-D plot of the conductance density as a function of the magnetic field in Tesla for the point $(k_x a, k_y a)$ = (2.40, 0.60) on the Brillouin zone shown in Fig.7. In order to obtain this plot the delta functions in the spectral weight have been expanded in the Fourier cosine series: $δ(x−a) = (2π)^{-1} + π^{-1}\sum_{m=1}^{∞} \cos[m(x −a)]$ . In this infinite series, 26 terms have been taken into account. The plot is more of illustrative nature than quantitative, as its origin is a single relevant point on the Brillouin zone. Obviously, a way needs to be found to extract contributions from all the relevant points and perform the **k**-integrations to obtain a near-comprehensive idea of the features of the oscillations. Other reasons for the plot being non-quantitative are (i) a direction is very much required as to how many terms in the δ(x−a)-series are needed to make the outcome a robust one, and (ii) the assumption – ($\partial f / \partial ε$) = δ(ε−μ) made above is, strictly speaking, valid for a metal with a spherical Fermi surface. We hope to resolve these issues in a future communication.